%% file: NN.tex
\documentclass[]{elsarticle}

\usepackage{amsmath}
\usepackage{amssymb}
\usepackage{graphicx, subfigure}
\usepackage{array,longtable}
\usepackage[english]{babel}
\usepackage{multirow}
\usepackage{xspace}
\usepackage{booktabs}
\usepackage[sc]{mathpazo}
\usepackage{mathtools}
\usepackage{xcolor}
\usepackage{csquotes}
\usepackage[hyphens]{url}

\newcommand{\ep}{EPOS-LHC }
\newcommand{\qg}{QGSJETII-04 }
\newcommand{\qgj}{QGSJETII-04}
\def\Offline{\mbox{$\overline{\textrm%
{Off}}$\hspace{.05em}\protect\raisebox{.4ex}%
{$\protect\underline{\textrm{line}}$}}\xspace}

\begin{document}
\thispagestyle{empty}

\title{Deep learning techniques applied to the physics of extensive air showers}

\author[a1]{A. Guill\'en}
\author[a2]{A.~Bueno}
\author[a2]{J. M. Carceller}
\author[a2]{J. C. Mart{\'\i}nez-Vel\'azquez}
\author[a2]{G. Rubio}
\author[a2,a3]{C. J. Todero Peixoto}
\author[a2]{P. Sanchez-Lucas\fnref{f1}}
\fntext[f1]{now at U. of Z\"urich}
\address[a1]{Dpto. de Arquitectura y Tecnolog{\'\i}a de Computadores, Universidad de Granada, Granada, Spain.} 
\address[a2]{Dpto. de F{\'\i}sica Te\'orica y del Cosmos \& C.A.F.P.E., Universidad de Granada, Granada.} 
\address[a3]{Depto. de Ci\^encias B\'asicas e Ambientais, Escola de Engenharia de Lorena, U. of S\~ao Paulo, S\~ao Paulo, Brazil.}
\begin{abstract}

Deep neural networks are a powerful technique that have found ample applications in several branches of physics. In this work, we apply deep neural networks to a specific problem of cosmic ray physics: the estimation of the muon content of extensive air showers when measured at the ground. As a working case, we explore the performance of a deep neural network applied to large sets of simulated signals recorded for the water-Cherenkov detectors of the Surface Detector of the Pierre Auger Observatory. The inner structure of the neural network is optimized through the use of genetic algorithms. To obtain a prediction of the recorded muon signal in each individual detector, we train neural networks with a mixed sample of simulated events that contain light, intermediate and heavy nuclei. When true and predicted signals are compared at detector level, the primary values of the Pearson correlation coefficients are above 95\%. The relative errors of the predicted muon signals are below 10\% and do not depend on the event energy, zenith angle, total signal size, distance range or the hadronic model used to generate the events.  
\end{abstract}
\maketitle

\section{Introduction} 
\label{sec:int}
Ultra-high-energy cosmic rays (UHECR) are atomic nuclei with energies above $10^{18}$ eV. They constitute the most energetic form of radiation ever detected by humankind. Note that, when expressed in a reference frame where one of the protons is at rest, the energy of the proton beam accelerated at LHC in its present configuration reaches $10^{17}$ eV. UHECR were discovered more than 50 years ago \cite{Linsley:1963km}. However, we still lack a plausible theory that explains the chemical composition of this radiation, where it is produced and what mechanisms are capable of conferring to these nuclei such extraordinary energies. We know that their flux is a steeply decreasing function of energy \cite{Verzi:2017hro}. For example, for energies above $10^{19}$ eV, we detect on average one particle per km$^2$ per year. This fact renders direct primary particle detection impractical. However, an alternative way to study the properties of UHECR exists, which is through the analysis of the extensive air showers produced after UHECR collide with the nuclei of the Earth's atmosphere \cite{Anchordoqui:2018qom}.     

The Pierre Auger Observatory is the biggest instrument built so far to scrutinize the mysteries of UHECR \cite{ThePierreAuger:2015rma}. It works in the so-called {\it hybrid mode}: it is able to measure, with fluorescence telescopes, the light produced by the de-excitation of the atmospheric nitrogen after the passage of the air-shower and, it samples as well the particles of the air-shower that reach the ground with an array of water-Cherenkov detectors, covering a sparse surface of 3000 km$^2$. The latter is known as the Surface Detector (SD). The signals recorded by the surface stations are triggered by a mix of two kinds of components: electromagnetic (photons, electrons and positrons) and (anti)muons. The construction of the Pierre Auger Observatory was completed in 2008, but it has been taking data steadily since 2004 for a total accumulated exposure that nowadays is above 10$^5$ km$^2$ sr year. This is the largest data sample ever collected in this field of physics to advance our understanding of the properties of the non-thermal Universe. 
 
The SD is not optimized to separate in a clean way the signals produced by the muons traversing it from those produced by the electromagnetic component. The measurement of the muon signal is of the utmost importance to understand how well models of hadronic interactions describe the data \cite{Aab:2016hkv,Aab:2014pza}. It is also key to perform an analysis of the mass composition of the primary flux of UHECR on an event-by-event basis. This kind of analysis might open the door to solving the mystery of where UHECR are produced and what is the origin of the observed flux suppression at extreme energies \cite{Aab:2017njo}.. The prospects for enhanced mass composition analyses and the observed mismatch in the number of muons between data and simulations are the main motivations behind the upgrade of the Auger Observatory, dubbed AugerPrime \cite{Aab:2016vlz}. 

Recently the Telescope Array collaboration has also reported an excess of events with large muon purity when compared to predictions from simulations \cite{Abbasi:2018fkz}. Therefore, given its importance, it comes as no surprise that the problem of estimating the muon content of extensive air showers at the ground had been tackled in several ways by different collaborations. For example, AugerPrime proposes a solution based on instrumentation: each water-Cherenkov detector is equipped with a complementary plastic scintillator detector on top of it. This new detector is mostly sensitive to the electromagnetic component of the extensive air-shower. Therefore it is through the combination of the two signals, simultaneously registered by each enhanced station, that the muon signal is obtained. The matrix formalism to be employed to extract the muon contribution is similar to the one proposed in \cite{Letessier-Selvon:2014sga}. 

Focusing on the particular case of the Pierre Auger Observatory, we discuss here a new complementary approach to estimate the amount of muonic signal. It is a software solution based on the use of deep neural networks \cite{dnn1,dnn2}. We have used Monte Carlo events originated by primaries of several species. They have been fully simulated and reconstructed using the official software of the Pierre Auger Collaboration \cite{Argiro:2007qg}. With them we asses the performance of this new method in the energy range where the Auger SD is fully efficient. 

\section{Deep Neural Networks: design, training and optimization}
\label{sec:deep}

\begin{figure*}[ht]
\includegraphics[width=0.49\linewidth]{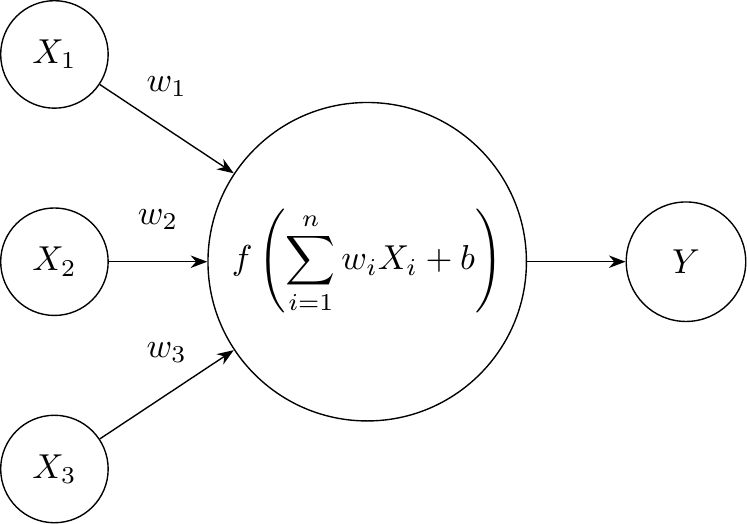}
\def\svgwidth{.49\linewidth}
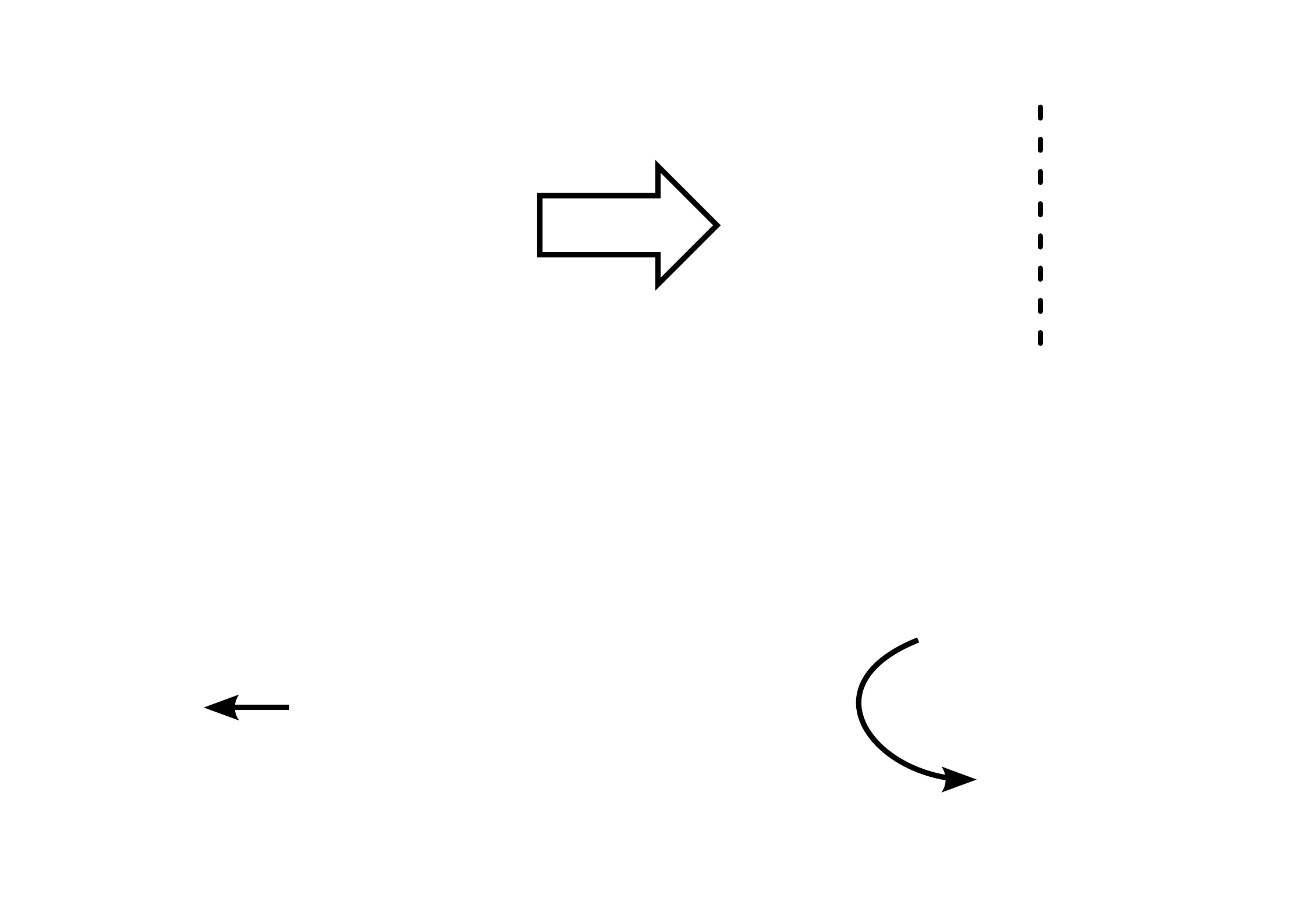
\caption{(Left) Example of a single neuron with three inputs: $X_i$. The three weights $w_j$ and the bias $b$ are free parameters and will be adjusted during the training process. The neuron computes an activation function $f$ that gives an output $Y$. (Right) Diagram of the genetic algorithm used in this work. Deep neural networks (DNN) are built with random numbers of hidden layers and neurons per layer and trained. The DNN with better performance are selected in binary tournaments. Only for the selected DNN, we cross the number of neurons in some layers between individuals and we introduce mutations (change randomly the numbers of neurons in certain layers). This process provides a new generation of DNN ready to be trained.}
\label{fig:diagrams}
\end{figure*}

Artificial neural networks are loosely based on biological neural networks. Both kind of networks are made of neurons or nodes that are interconnected, with information being transmitted through
these connections. In artificial neural networks 
each neuron computes a function of a linear combination of its inputs: the \emph{activation function}.
The input of this function can be the data that we feed the neural net with
or the output of other neurons. The output of this function can then be used
by other neurons. See the left panel of Figure \ref{fig:diagrams}.

Neurons are grouped in layers, with the first layer being the input layer,
the last one being the output layer and those in the middle are called hidden layers. Layers are fully connected. 
In this work we use Deep Neural Networks (DNN), i.e. with several hidden layers, and supervised learning. Therefore, 
in the first stage, the DNN is trained on labelled data:
the predictions of the DNN for the value of the labelled variable and the true values of the labelled variable are compared, and the error 
is minimized. This is done by updating the weights and biases in the linear combinations
that make up the input of each neuron by utilizing back-propagation algorithms.  

In the following subsections,
the technical aspects of the DNN that we have built are explained.
There are several possible choices when building a DNN:
the number of neurons in each layer, the number of layers,
the activation function, etc.
We explain the process followed to chose these hyperparameters, and in the end,
we give an overview of the final architecture of the DNN that we used. The code has been implemented using the functions provided by the SciPy ecosystem \cite{SciPy} (Numpy and Matplotlib), Scikit-learn \cite{scikit-learn} and Keras \cite{chollet2015keras} using Tensorflow \cite{tensorflow}, all interpreted by Python (version 3.6).

\subsection{Data preprocessing}

Following the recommendations in \cite{Goodfellow-et-al-2016}, the data used as input and the output (muon signal) have been normalized each one independently with mean \(\mu=0\) and standard deviation \(\sigma=1\). Before normalizing, some variables have been transformed to help the DNN learn. The details are explained in Section \ref{sec:mc}.

\subsection{Activation function and weight initialization}
\label{sec:activation}
Each neuron computes a function of the linear combination of its inputs, and its output is connected to each neuron in the subsequent layer. There is a wide variety of functions that can be chosen, although, following the recommendations in \cite{Nair-Hinton-2010,Glorot-2011a}, the Rectifier Linear unit (ReLu, defined as \(f(x) = \max \{ 0 , x \}\)) could be the best choice. Nonetheless, instead of choosing it ourselves, we have let the genetic algorithm chose the one that gave the best performance among the ones we tested (linear, tanh, softmax, ReLu and self-normalizing exponential linear units, SELU).

Regarding the initialization of the weights of the neural network, after analysing two methods (setting all weights to zero or random initialization), we have obtained the bests results using a random weight initialization that takes values from a normal distribution with mean \(\mu=0\) and standard deviation \(\sigma=0.05\). 

\subsection{Optimization algorithm and loss function}

During the training process, the weights in the connections between each layer have to be optimized. To do so, it is necessary to define a loss function, that is, a function that measures how well the neural net is predicting its output. As we are dealing with a regression problem and no other restrictions are known, one reasonable criterion to determine if the model has a good performance is the mean squared error (MSE):
\begin{equation} \label{eq:1}
\mathrm{MSE}=\frac{1}{n}\sum_{i=1}^n(Y_\mathrm{true}^{(i)}-Y_\mathrm{pred}^{(i)})^2 
\end{equation}
where \(n\) is the number of samples, \(Y_\mathrm{true}^{(i)}\) is the true value that we want
to predict for the $i$-th example and \(Y_\mathrm{pred}^{(i)}\) is the prediction done by the neural network for the same example.
This is the function that will be minimized during the training process. In our case, the variable $Y$ corresponds to the total muon signal recorded in each individual water-Cherenkov detector.

The optimization algorithm used was the Adam (Adaptative momentum) \cite{Adam}, which is a stochastic gradient-based method. It is the recommended algorithm in many deep learning applications. We followed the common usage of running the algorithm using the default parameters recommended in \cite{Adam}. The gradients are computed and then corrected using a first and second raw moment estimate, leading to the new value of the parameters to be optimized.

\subsection{Genetic algorithm}

Genetic algorithms (GAs) are global optimization algorithms that have been widely used in many areas where there is no better way of doing things than trying random numbers \cite{Mitchell-Melanie-1996}. The idea beneath these algorithms is to imitate the behaviour of nature by evolving populations of individuals through time, in a way that only the characteristics (or genes) of the fittest individuals are propagated into the next generation of the population, see the right panel of Figure \ref{fig:diagrams}. We have let the width (number of neurons in each layer), depth (number of hidden layers) and choice of activation function free for the genetic algorithm to pick the best combination.

The first step is to decide what to choose as the individual that will be evolved over time. For the sake of simplicity, we have defined our DNN as a vector of a certain length, whose elements are a natural number in the interval \([0,100]\). This set of natural numbers corresponds to the number of neurons in each layer. This vector also has an integer that maps the type of activation function (see list in \ref{sec:activation}), and therefore, the algorithm can check different combinations of them. The number of hidden layers was limited to a maximum of ten.

We generated randomly 50 individuals, or 50 neural networks with random number of neurons in each layer, random number of layers and random activation functions. We have repeated the following process for 100 generations. To compute the fitness of each individual, networks are trained over ten epochs to have an estimation of their potential performance. The validation is done over a sample independent of the one used for training. Once they are evaluated, a binary tournament selection is carried out to obtain the subset of individuals that are the ancestors for the generation of the offspring. Then there is a two-point crossover, that is, the number of neurons in each layer from a certain start layer \(l_0\) up to an end layer \(l_1\) are exchanged between two individuals with probability 0.8. The last step is to mutate or change randomly, with probability 0.1, the number of neurons in each layer of an individual. When the new population is available, we have used an elitism mechanism where the best individual in the current population is included into the next one. All these values were set up according to the literature, following the same design principles discussed in \cite{GONZALEZ200732, GUILLEN20093541, book1, article1} and after checking by some experiments that higher values did not produce a significant improvement. 

A filter was also applied. When any layer was found to have less than five neurons, it was discarded. In this way, we make sure that the layer is really needed and there is no need to ``dropout'' or apply regularisation afterwards. The last layer only has one neuron and is used to obtain the output. 

\subsection{Final DNN structure}

\begin{figure*}[ht]
        \centering
	\includegraphics[scale=0.4]{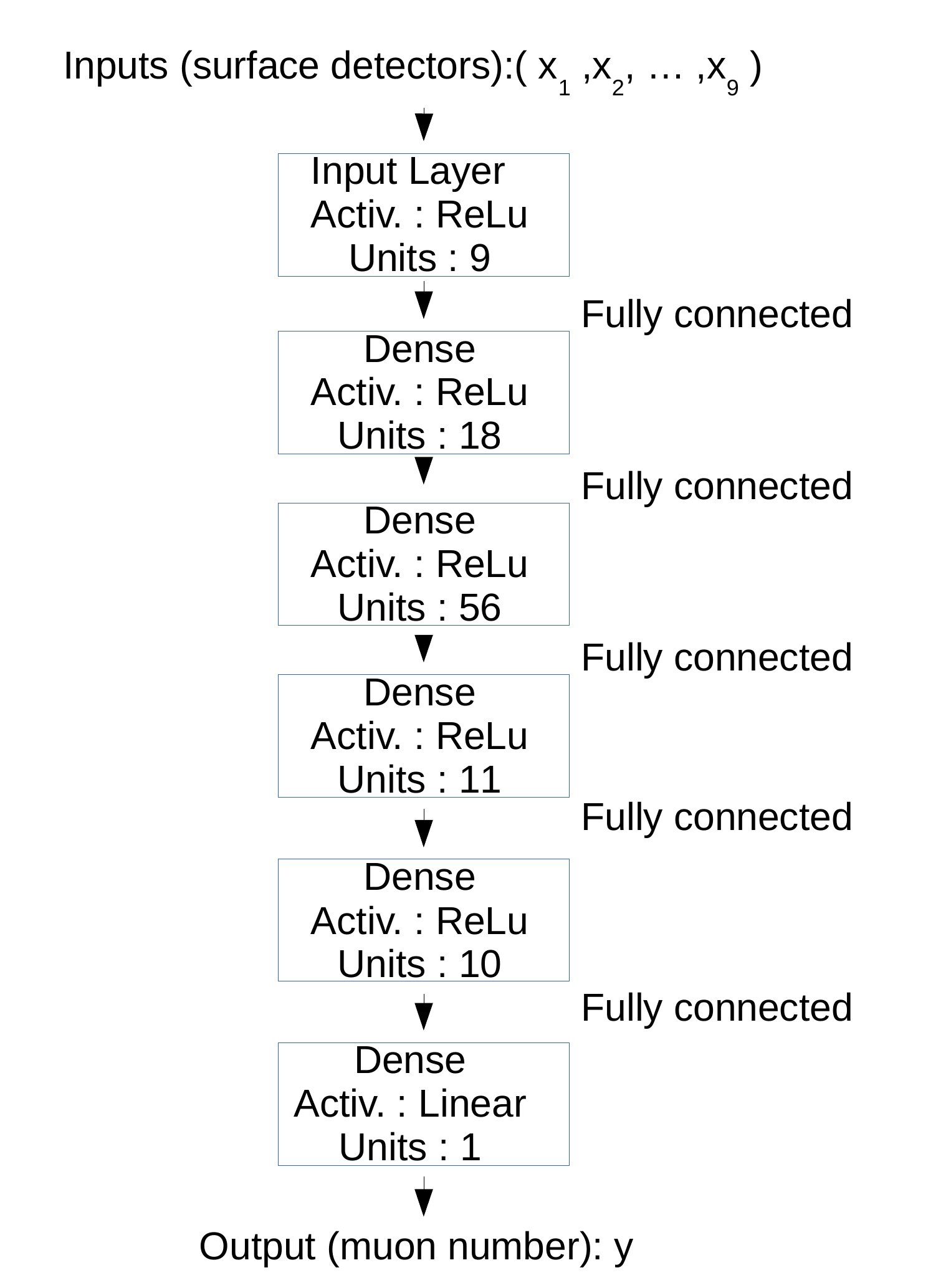}
	\caption{Final structure of the DNN coming from the GA optimization. It has five fully connected layers using ReLu as activation function. The numbers of neurons in each layer are indicated.}
	\label{fig:finalDNN}
\end{figure*}

The final DNN obtained is represented in Figure \ref{fig:finalDNN}. It is the outcome of running over a mix of equal fractions of proton, helium, nitrogen and iron nuclei generated with \qg that are independent of those used in section \ref{sec:mc}. The network is made up of six layers: five fully connected layers using ReLu as activation function and a final layer that combines the outputs from the fifth layer linearly. In this neural net, complexity starts increasing from low to high (from 9 to 56 neurons) and then it decreases drastically to the point where it started. The first runs of the GA, using a maximum of ten layers, always decreased the number of them to six or seven, so to simplify the structure of the DNN we fixed the number of layers to six, without a loss of performance. So we executed the GA again (with a maximum number of layers set to six) obtaining the current configuration. In this way, once we explored complex solutions regarding the network depth, we exploited the reduced solution space to determine more precisely how many neurons per layer should be included.  

\section{Application of DNN to the problem of measuring the muon signal recorded by the water-Cherenkov detectors of the Pierre Auger Observatory}
\label{sec:mc}
\begin{figure*}[ht]
	\centering
	\includegraphics[scale=0.4]{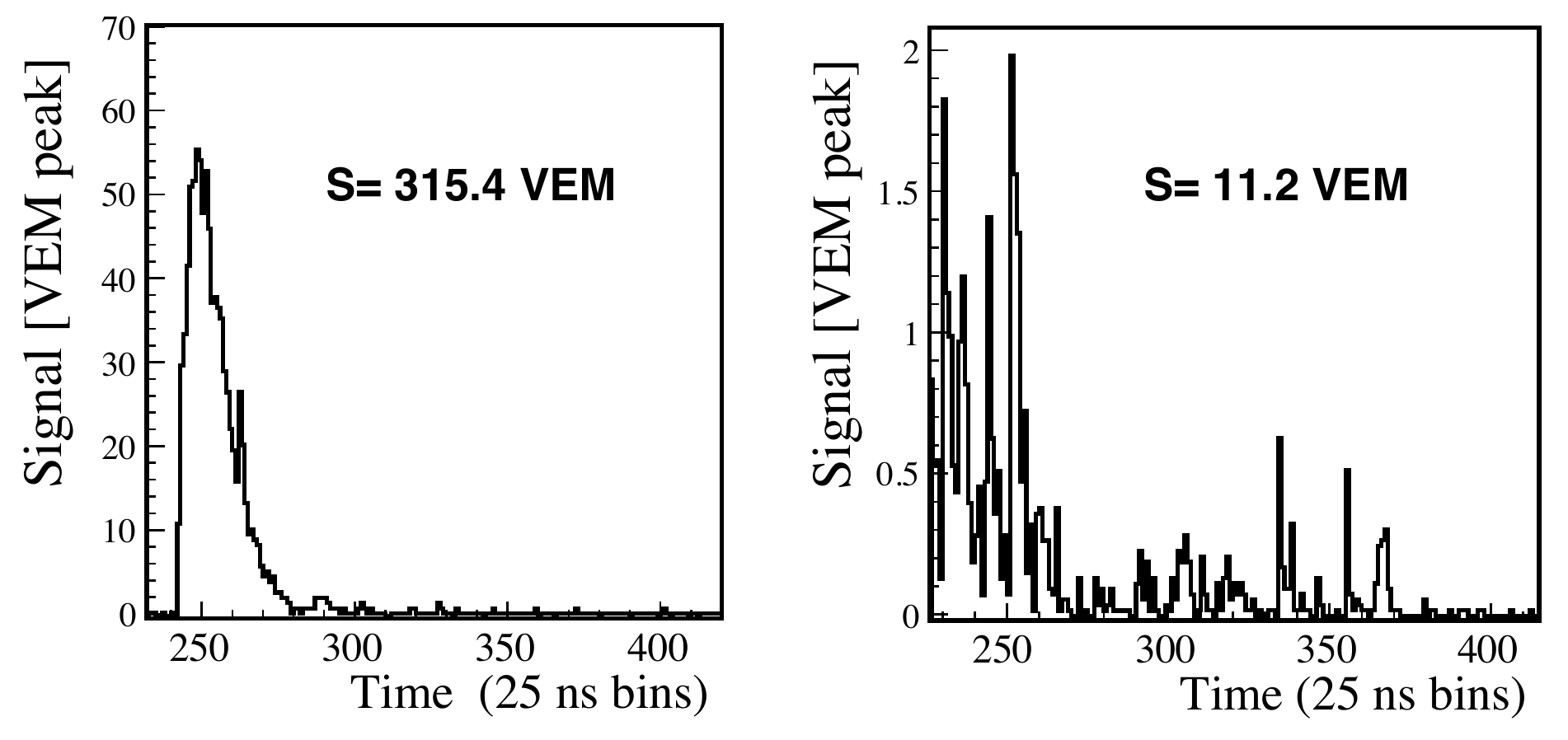}
	\caption{FADC traces corresponding to triggered detectors located at 650 m (left) and 1780 m (right) from the reconstructed core of an extensive air-shower. The traces, corresponding to a typical vertical event of $3\times 10^{19}$ eV, are mix of electromagnetic and muonic particles. The spiky structure is attributed to be caused mainly by muons, while the accumulations of signals left by photons or electrons/positrons are usually more spread in time. These traces are reproduced from \cite{ThePierreAuger:2015rma}.}
	\label{fig:FADC}
\end{figure*}

The Surface Array of the Auger Observatory consists of 1660 water-Cherenkov detectors laid out in a triangular grid, whose separation between closest-neighbours is 1500 m. These detectors provide the arrival times of the particles that impinge on them. Each detector is equipped with three PMTs, the signals registered by them are digitized by 40 MHz 10-bit flash analog to digital  converters (FADCs). For each triggered station, 19.2 $\mu$s (768 bins of 25 ns each) of data are recorded by every FADC. The signals are expressed in units of VEM \cite{Bertou:2005ze}. A VEM (\enquote{vertical-equivalent muon}) is the total signal left by a muon that traverses the water-Cherenkov detector vertically at its center. Figure \ref{fig:FADC} shows two examples of typical FADC traces. 

As stated before, the FADC traces are a mix of the electromagnetic and muonic components present in extensive air showers. To fully exploit the physics information encoded in the signals recorded by the water-Cherenkov detectors, it is a must to separate the contributions of the different particle species that traverse the stations. Therefore, the goal of this study is to estimate, for every single triggered detector used in the reconstruction of an event, the amount of signal that corresponds to the energy deposited by muons.  

We feed the neural network with the set of variables described below. Notice that some of them pertain to the global features of the event (items 1 and 2 in the list below), others refer specifically to the information furnished at station level. The variables chosen to provide information at station level are those customarily used in the physics analyses of the Pierre Auger collaboration \cite{Aab:2017njo}. The total trace used in our analysis is computed as the average of the traces recorded by each of the active PMTs of a station. In our analysis, we do not use detectors with signs of saturated traces. We disregard as well stations whose measured total signal is below 10 VEM (above this value the probability of single detector triggering is close to 100\%). We do not work with events whose energy lies below $3\times 10^{18}$ eV since this is the minimum energy at which the Auger trigger efficiency reaches 100\% for the events recorded by the 1500 m array. In this way, we avoid applying correction factors associated with trigger inefficiencies. The event selection efficiency is above 95\% for the energy interval of interest ($\geq 3\times 10^{18}$ eV). The chosen list of variables read as follows:

\begin{enumerate}
\item True Monte Carlo energy. This is the only Monte Carlo variable used. Since this study deals only with simulated events, we discarded to use reconstructed energies to avoid all the discussion associated to the energy calibration of SD events. We use as input variable the decimal logarithm of energy, which is expressed in eV.
\item Zenith angle: $\theta$. It is measured in degrees and provides the incoming direction of the cosmic ray. For this study and due to the lack of simulated horizontal events, we limit ourselves to angles below 45 degrees\footnote{In principle, we cannot think of any showstopper that prevents us from applying the same kind of treatment to more horizontal events.}. 
\item Distance of the station to the reconstructed core of the air-shower. We measure it in meters. 
\item Total signal registered by the station. We measure it in VEMs.
\item Trace length. Number of 25 ns bins between the Start and End bins of the trace \cite{Argiro:2007qg}. 
\item Polar angle of the detector with respect to the direction of the shower axis projected on to the ground. It is measured in radians.
\item Risetime, $t_{1/2}$ (computed as discussed in \cite{Aab:2017cgk}). The risetime is defined as the time it takes for the integrated signal to grow from 10\% up to 50\% of its total value and it is measured in ns. 
\item Falltime (computed following \cite{Aab:2017cgk}). Time for the integrated signal to go from 50\% to 90\% of its total value and it is measured in ns.
\item Area over peak. It is defined as the ratio of the integral of the FADC trace to its peak value.

\end{enumerate}

We note that the true value of the total muonic signal $S^{\mathrm{true}}_{\mu}$ (measured in VEMs) is the output value of the DNN. The evaluation metric used to gauge the performance of the DNN is the MSE defined in Equation \ref{eq:1}. 

The set of mixed global and local (station level) variables, listed above, performs well. Enlarging it with new variables to obtain a better estimation of the muon signal just represents an increase of the training time but does not improve our results. In particular, one can think that feeding the net with the whole temporal series of bins that form the trace would substantially improve the precision with which the muon signal is extracted. We observed that in that case the gain is almost negligible, while the computational cost in terms of time increases by a sizeable factor. We understood that to fully exploit the time information contained in the recorded traces we have to consider new classes of artificial neural networks. For example, an interesting possibility that we will explore in a future work is the use of recurrent neural networks to try and extract not only the total muon signal but also its time structure. 

 We decided to train the network with events generated using the \qg model  \cite{qgsjet04}. The events have been generated with a distribution that is uniform in energy. Once the internal network architecture is fixed and its performance evaluated with \qg events, we use \ep \cite{epos} to show how our results depend on the different assumptions made to model hadronic interactions at ultra-high energies. Another vital decision is related to the nature of the nucleus (or set of nuclei) used to train the neural network. We observed that training the network with a single species is a far from optimal decision. Table \ref{tab:dnn} shows the values of the MSE (our evaluation metrics) and, the mean values of the distributions obtained as the difference between the true and the predicted muonic signals. For example, when iron nuclei are used to build the network, the number of predicted muons in events generated by protons is overestimated. 
 Similar conclusions are drawn when a pure sample of protons is used to train the network and, later, we assess its performance on iron nuclei, in this case we underestimate the muonic signals produced by Fe. Since protons and iron nuclei differ in the values of the local variables for fixed values of the energy and zenith angle, the use of a single species to define the internal structure of the network does not guarantee an efficient coverage of the full available phase space of possible configurations. The situation improves when the network is trained with a mix of iron nuclei and protons in equal amounts. However, we observed that our estimations of the muonic signal improve even more when a mix of equal fractions of proton, helium, nitrogen and iron nuclei is used in the final training sample. As shown in Table \ref{tab:dnn}, this is the combination that offers the best performance. 
 
 \begin{table}[ht]
\centering
\caption{Selection of the set of simulated nuclei used to train the DNN. We use \qg as the model to simulate hadronic interactions. The third and fourth columns show the mean of the distributions of true minus predicted muon signals, measured in VEM, for proton and iron nuclei, respectively.}
\label{tab:dnn}
\begin{tabular}{|c|c|c|c|}
\hline
Training sample & MSE & Mean (p, VEM) & Mean (Fe, VEM) \\ \hline
Pure p & 6.8 & $-$0.16 & 0.26 \\ \hline
Pure Fe & 7.5 & $-$0.33 & 0.18 \\ \hline
Mix 25\% (p, He, N, Fe) & 6.4 & $-$0.08 & 0.16 \\ 
\hline
\end{tabular}
\end{table}   
 
 As a recapitulation, Table \ref{tab:events} shows the numbers of events generated and the stations used for two models of hadronic interactions. \qg events are used to train, test and validate the deep neural networks algorithms. \ep events are used for testing purposes only. The validation sample is used to choose the model that works better and also to study whether the learning process shows signs of overfitting, something that does not occur in the case under study. The events have been simulated using the CORSIKA package version 74004 \cite{corsika} and reconstructed using an official version of \Offline (i.e., the Auger simulation and reconstruction software \cite{Argiro:2007qg}). 
 
\begin{table}[ht]
\centering
\caption{Summary of the number of simulated events and detectors used in this work. Notice that the batches of stations used for training, validation and test correspond to distinct sets of detectors.}
\label{tab:events}
\begin{tabular}{|l|r|r|r|r|}
\hline
\multicolumn{5}{|c|}{\qg}                                                                                                                               \\ \hline\hline
\multicolumn{2}{|l|}{}                                            & \multicolumn{1}{c|}{Training} & \multicolumn{1}{c|}{Validation} & \multicolumn{1}{c|}{Test} \\ \hline
\multicolumn{1}{|c|}{Primary} & \multicolumn{1}{c|}{\# of events} & \multicolumn{3}{c|}{\# of detectors}                                                        \\ \hline
Proton                        & 19362                             & 16088                         & 4022                     & 57522                           \\ \hline
Helium                        & 12341                             & 15960                         & 3989                     & 34314                           \\ \hline
Nitrogen                      & 12201                             & 16071                         & 4017                     & 33739                           \\ \hline
Iron                          & 19478                             & 16076                         & 4018                     & 65231                          \\ \hline\hline
\multicolumn{5}{|c|}{\ep}                                                                                                                               \\ \hline\hline
\multicolumn{2}{|l|}{}                                            & \multicolumn{1}{c|}{Training} & \multicolumn{1}{c|}{Validation} & \multicolumn{1}{c|}{Test} \\ \hline
\multicolumn{1}{|c|}{Primary} & \multicolumn{1}{c|}{\# of events} & \multicolumn{3}{c|}{\# of detectors}                                                        \\ \hline
Proton                        & 18456                             & $-$                       & $-$                      & 78063                           \\ \hline
Iron                          & 18779                             & $-$                         & $-$                     & 86862                           \\ \hline\end{tabular}
\end{table} 
 
\section{Results}
\label{sec:res}

Once the approach described in the previous sections was optimized to extract an estimation of the muon signal recorded by the water-Cherenkov stations, we show to it fresh samples of simulated events generated with \qg and \ep models. Before discussing the outcome of this procedure, we note that this method can be readily applied to experimental data. The event selection efficiency is, as for the case of simulated events, close to one and we observed that the performance of the method is similar when the reconstructed energy is used. The only problem comes when interpreting the results, given the known inconsistencies between data and simulations \cite{Aab:2017cgk}. But this is a discussion that goes beyond the scope of this paper. 

In what follows, we use the following conventions in the ensuing plots: proton primaries are represented by red solid circles; He by brown open circles; N corresponds to orange open triangles and finally Fe is represented as blue solid squares.  

\subsection{\qg simulations: results at detector level}
\label{sec:qgdet}
\begin{figure*}[ht]
\centering
\includegraphics[width=\textwidth]{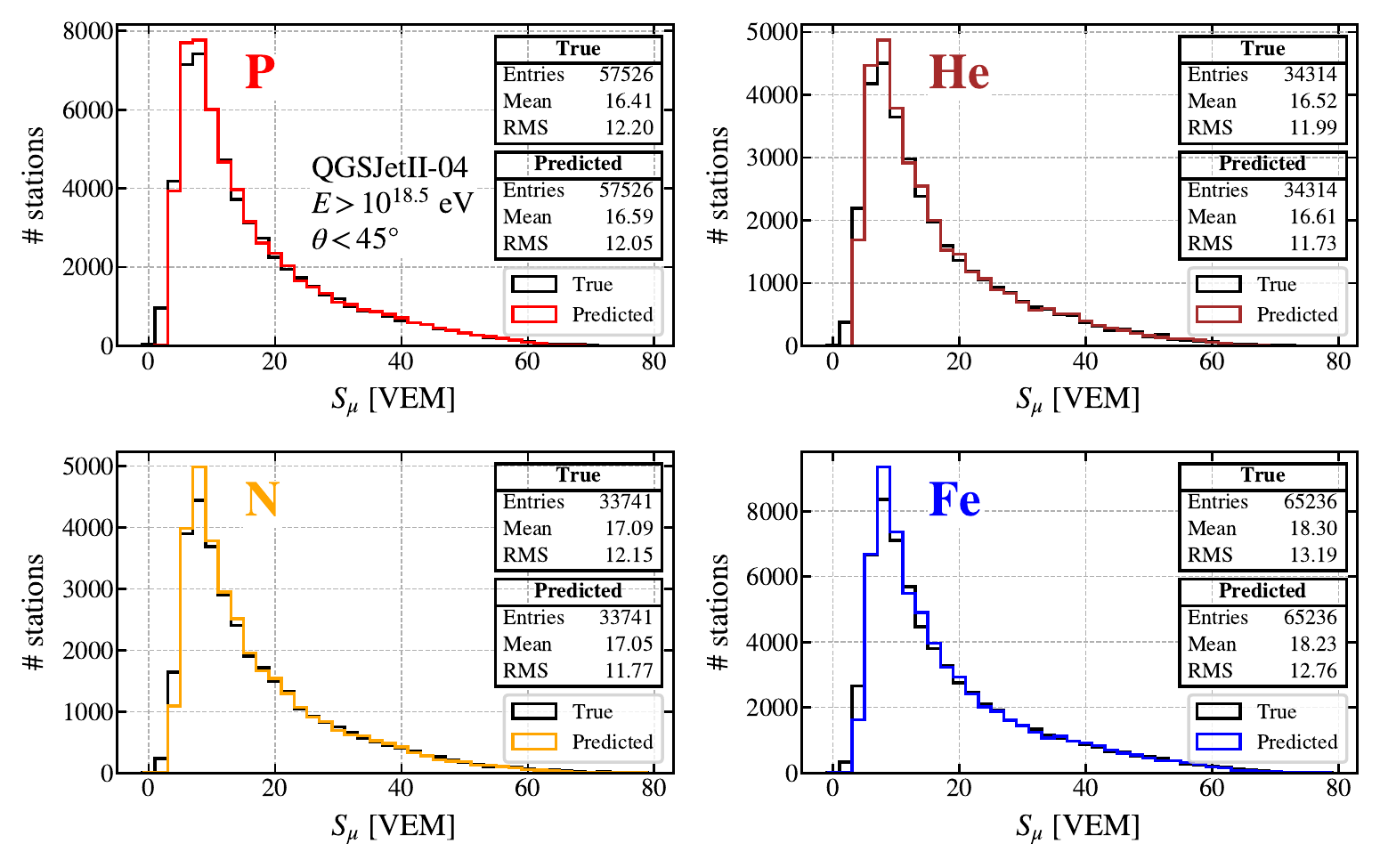}
\caption{(True and predicted muon signals for four different species of primary nuclei. Each entry corresponds to  the muonic trace recorded by each individual water-Cherenkov detector. The events have been generated in an energy interval that spans from log$_{10}$($E$/eV)=18.5 up to log$_{10}$($E$/eV)=20. Simulations use \qg to model hadronic interactions.}
\label{fig:Sm}
\end{figure*}

\begin{figure*}[ht]
\centering
\includegraphics[width=\textwidth]{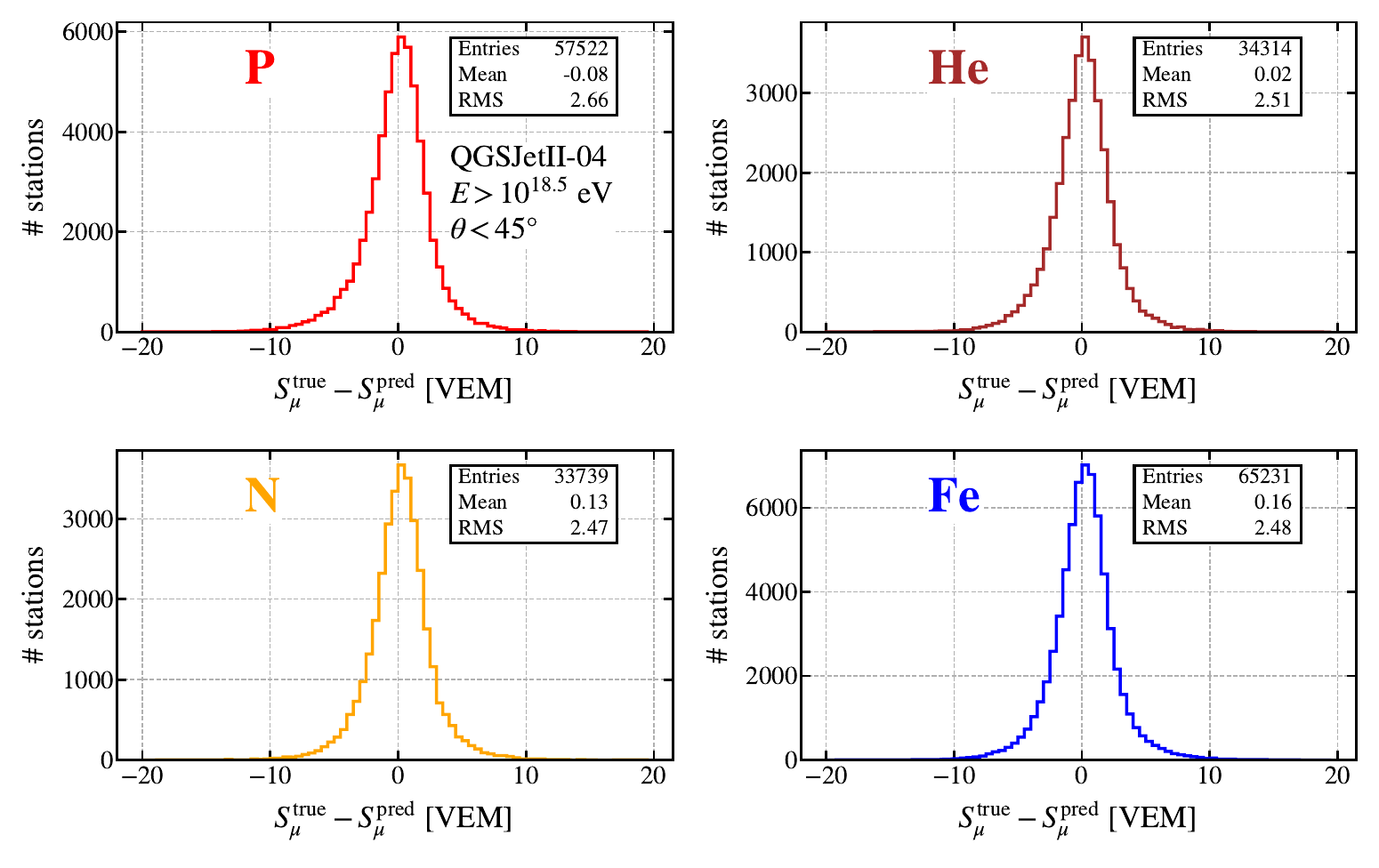}
\caption{Difference between true and predicted muon signals for four different kinds of primaries. Every entry corresponds to the information provided by a single water-Cherenkov detector. Events have been generated using \qg to model hadronic interactions.}
\label{fig:Smdif}
\end{figure*}

\begin{figure*}[ht]
\centering
\includegraphics[width=\textwidth]{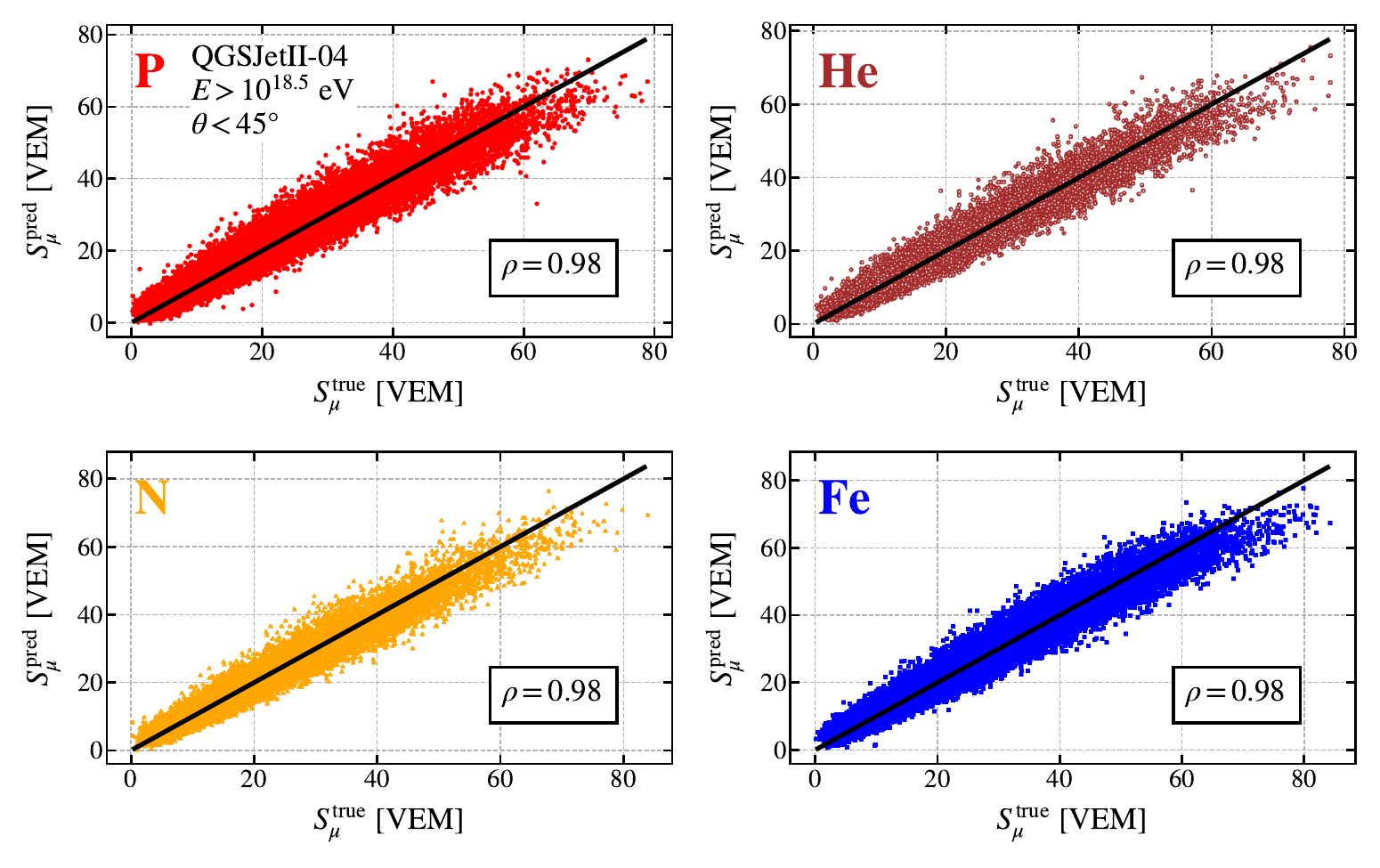}
\caption{Correlation between the true and the predicted muon signals. Events have been generated using \qg to model hadronic interactions.}
\label{fig:Smcorr}
\end{figure*}

Figure \ref{fig:Sm} shows, in units of VEM, the distribution of muonic signals for a set of events generated with \qgj. They have energies higher than $10^{18.5}$ eV and zenith angles up to 45 degrees. We consider stations with total measured signals above 10 VEM. The figures in this section show results at the single station level. For each species, we find that the distributions of predicted signals reproduce reasonable well the true signal distributions. This is illustrated in Figure \ref{fig:Smdif} where the difference between true and predicted signals is plotted for every nuclei. We obtain Gaussian distributions with means very close to zero and standard deviations around to 2.5 VEM. The accuracy in the prediction of the muonic signal is shown, this time as a scatter plot, in Figure \ref{fig:Smcorr}. The Pearson correlation coefficient is 0.98 for p, He, N and Fe. 


\begin{figure*}[ht]
\centering
\includegraphics[width=\textwidth]{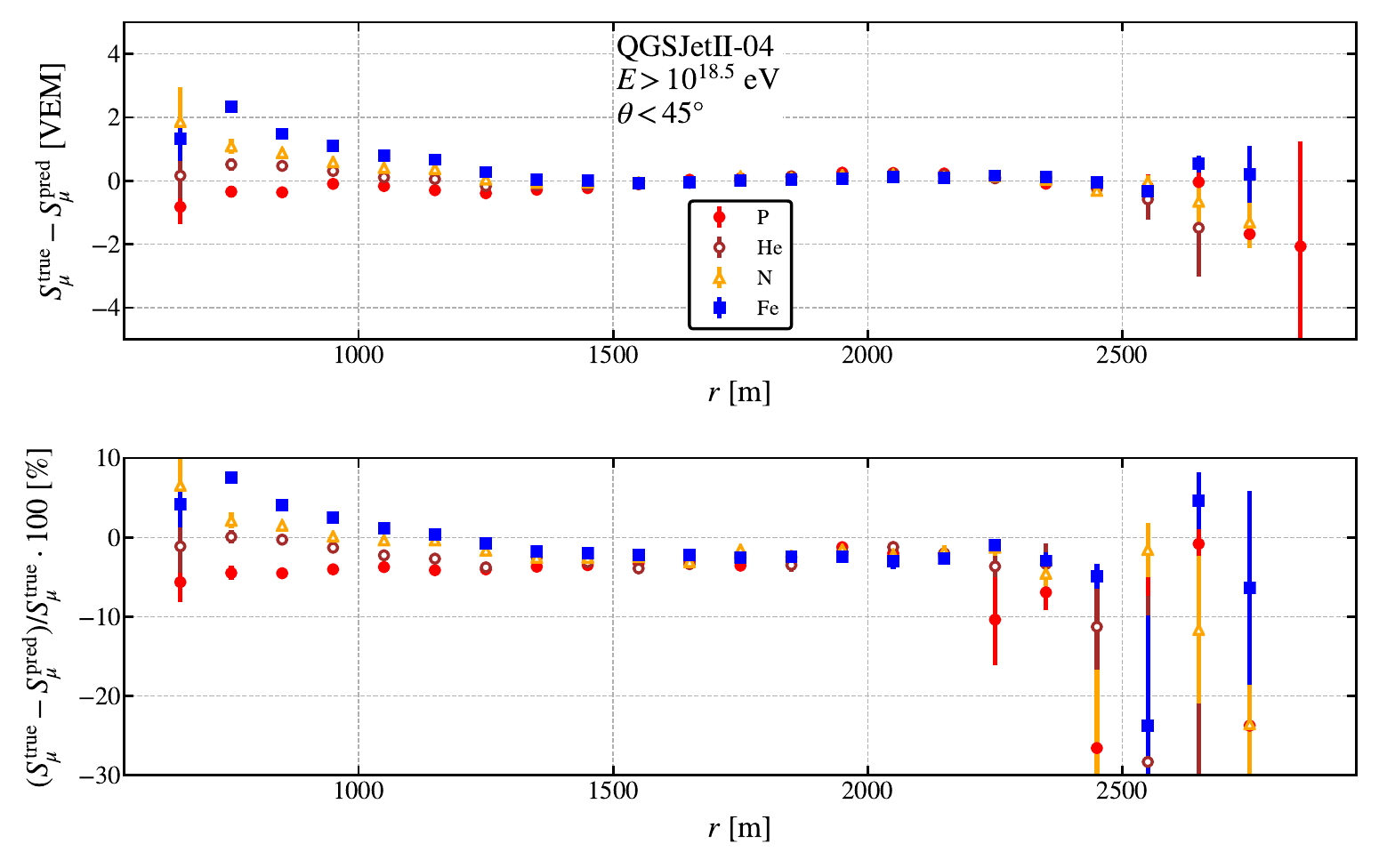}
\caption{Mean of the distribution of differences between true and predicted muon signals as a function of the distance to the core (top panel) and its associated relative error (bottom panel). Events have been generated using \qg to model hadronic interactions.}
\label{fig:Smr}
\end{figure*}

\begin{figure*}[ht]
\centering
\includegraphics[width=\textwidth]{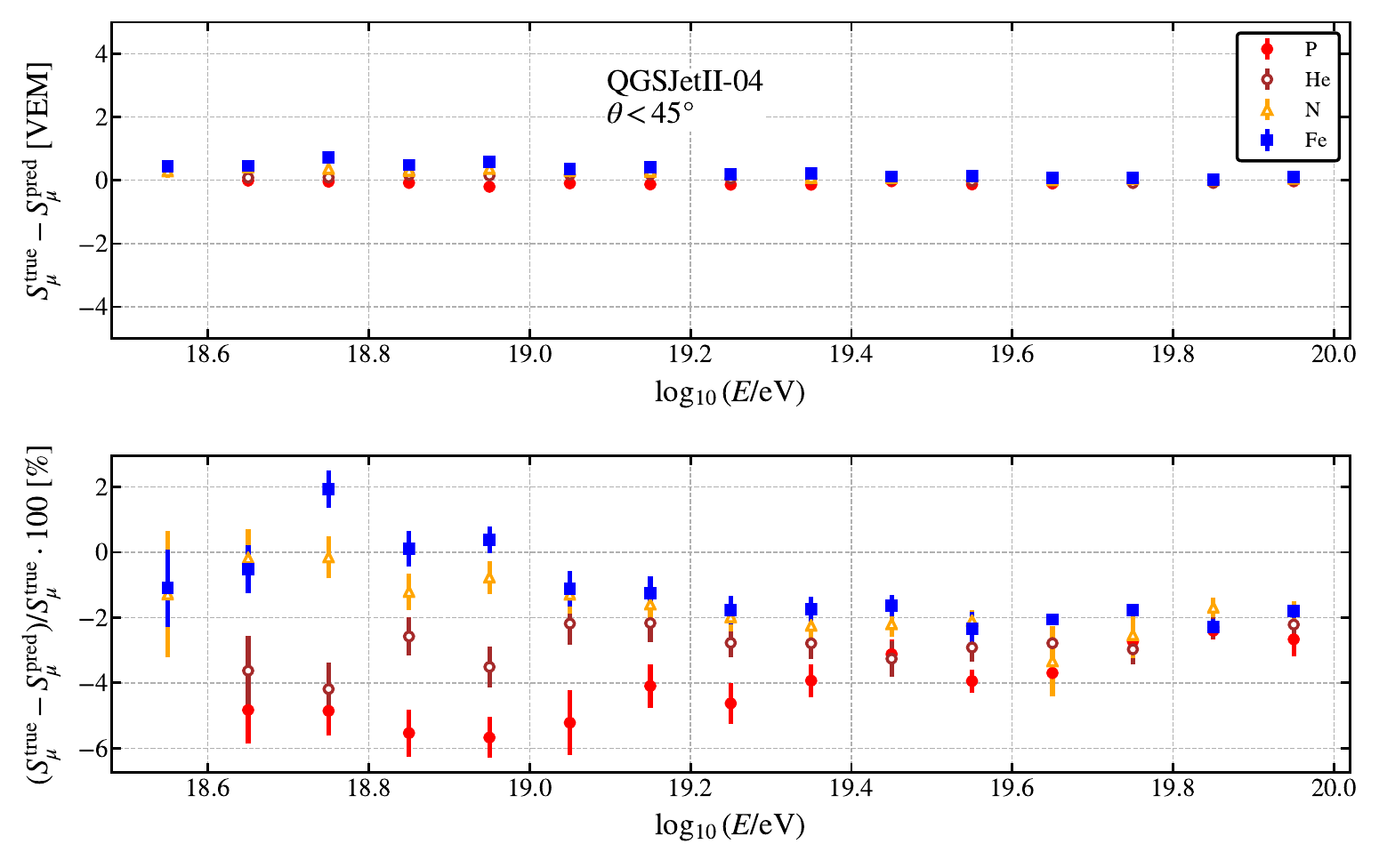}
\caption{Mean of the distribution of differences between true and predicted muon signals as a function of the event Monte Carlo energy (top panel) and its associated relative error (bottom panel). Events have been generated using \qg to model hadronic interactions.}
\label{fig:Sme}
\end{figure*}

\begin{figure*}[ht]
\centering
\includegraphics[width=\textwidth]{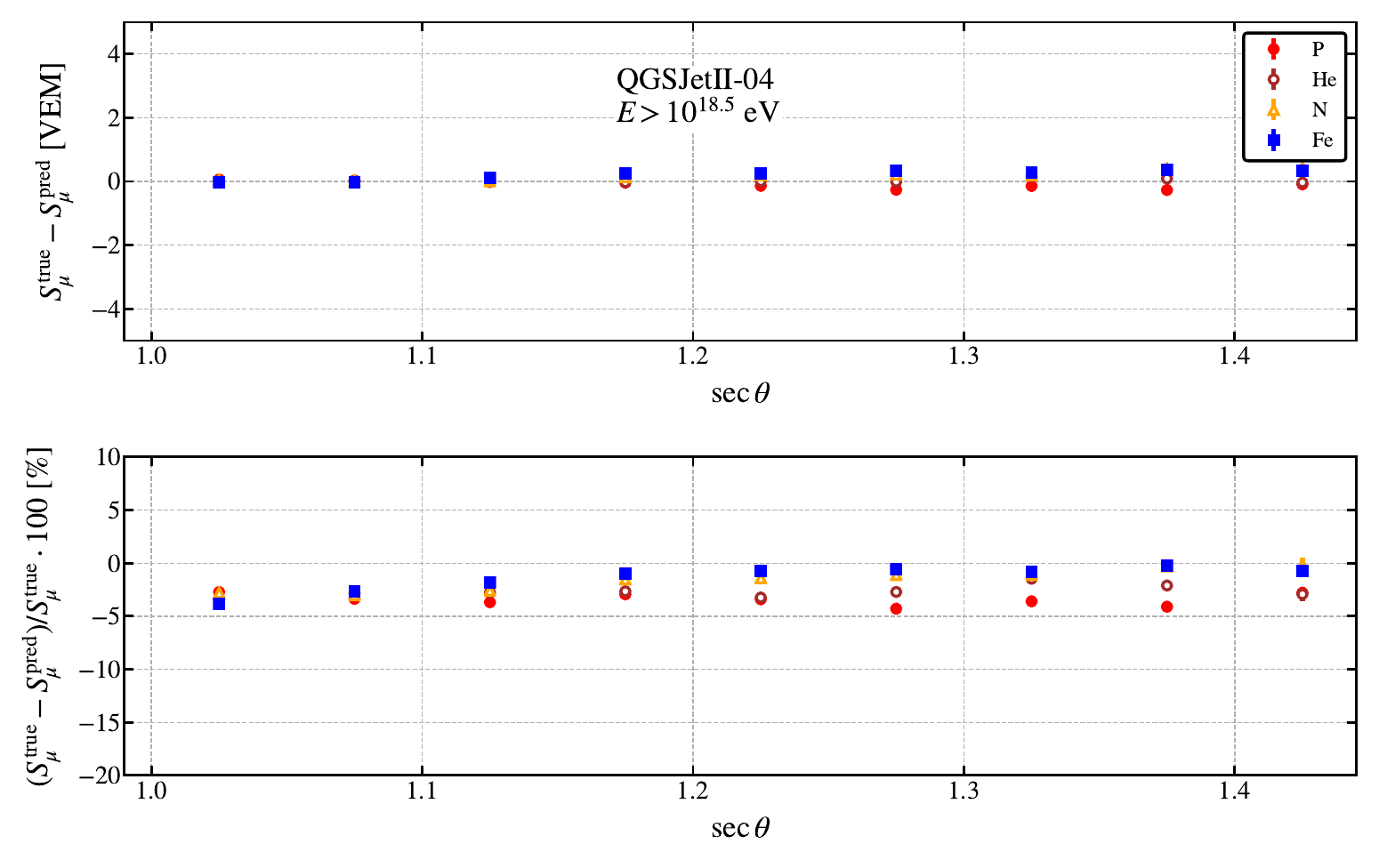}
\caption{Mean of the distribution of differences between true and predicted muon signals as a function of $\sec\theta$ (top panel) and its associated relative error (bottom panel). Events have been generated using \qg to model hadronic interactions.}
\label{fig:Smt}
\end{figure*}

\begin{figure*}[ht]
\centering
\includegraphics[width=\textwidth]{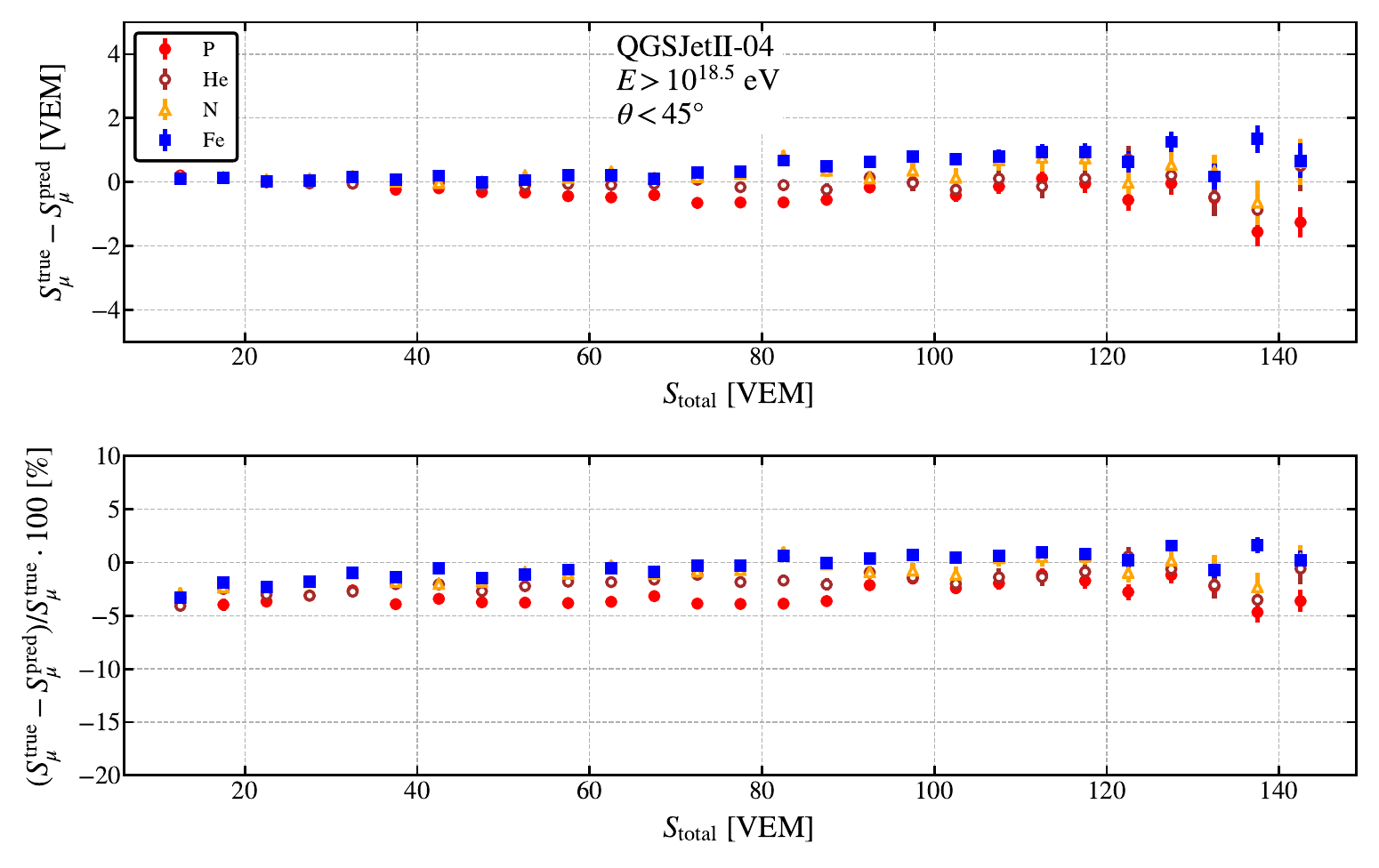}
\caption{Mean of the distribution of differences between true and predicted muon signals as a function of the total signal registered in individual stations (top panel) and its associated relative error (bottom panel). Events have been generated using \qg to model hadronic interactions.}
\label{fig:SmSt}
\end{figure*}

We have checked whether potential biases arise in the estimation of the muonic signal as a function of the following variables: distance of the station to the position of the core at the ground (Figure \ref{fig:Smr}), simulated energy of the event (Figure \ref{fig:Sme}), $\sec\theta$ (Figure \ref{fig:Smt}) and the total signal recorded by every water-Cherenkov detector (Figure \ref{fig:SmSt}). For this set of variables, the mean of the differences (in absolute value) between true and predicted signals are most of the time below 2 VEM. Relative errors are typically below 10\%. Our predictive power does not depend on the energy or the zenith angle of the air shower since the differences between predicted and measured signals are flat as a function of those two variables. At distances close to the core the spread in differences is wider. In addition to the fact that the number of events is smaller at short and very large distances to the shower core, we attribute part of this behaviour to the presence of a stronger contribution of the electromagnetic component. 

\subsection{\ep simulations: results at detector level}
\label{sec:epdet}

\begin{figure*}[ht]
\centering
\includegraphics[width=\textwidth]{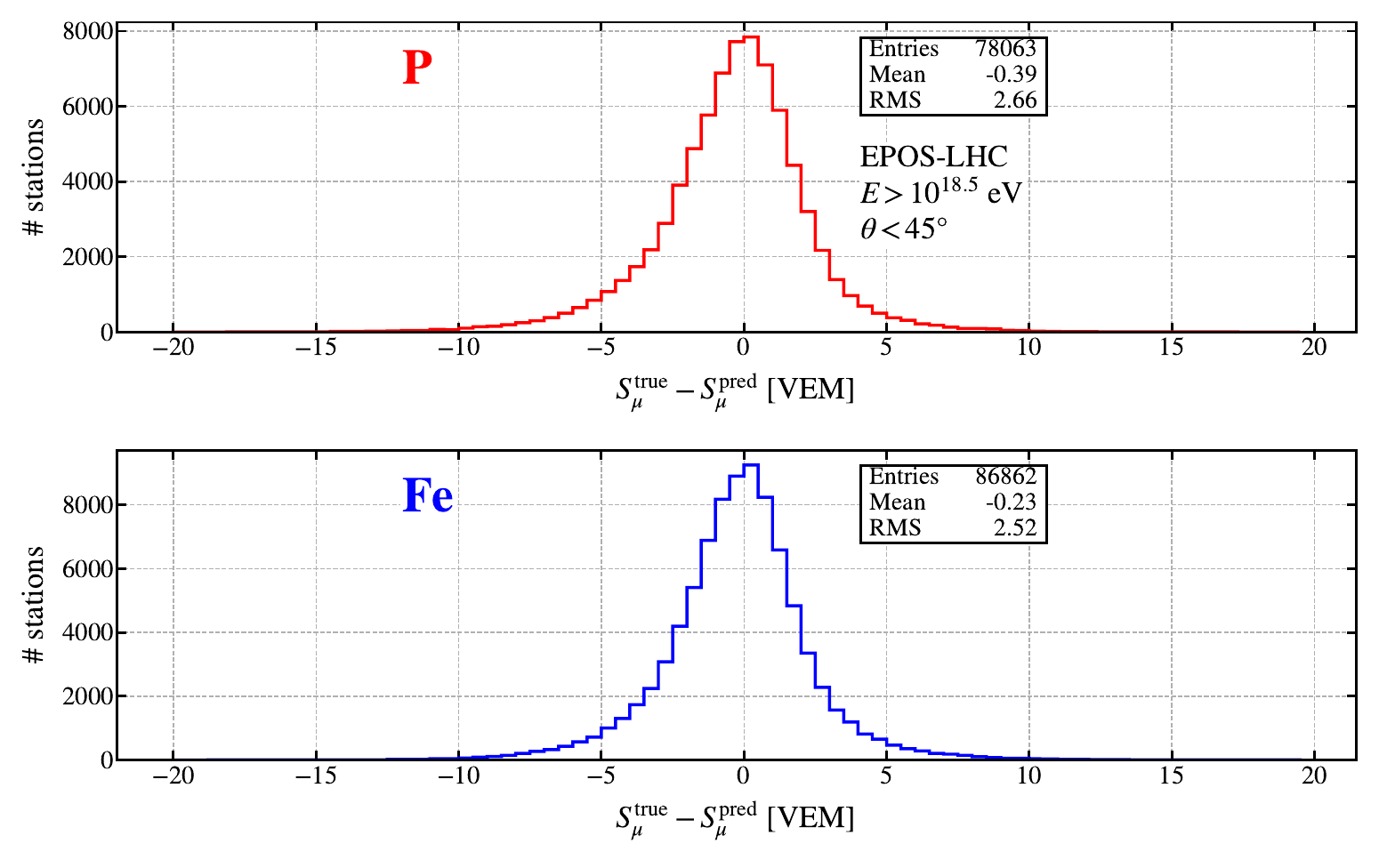}
\caption{Difference between true and predicted muonic signals at detector level for two different kinds of primaries. Events have been generated using \ep to model hadronic interactions.}
\label{fig:ESmd}
\end{figure*}

\begin{figure*}[ht]
\centering
\includegraphics[width=\textwidth]{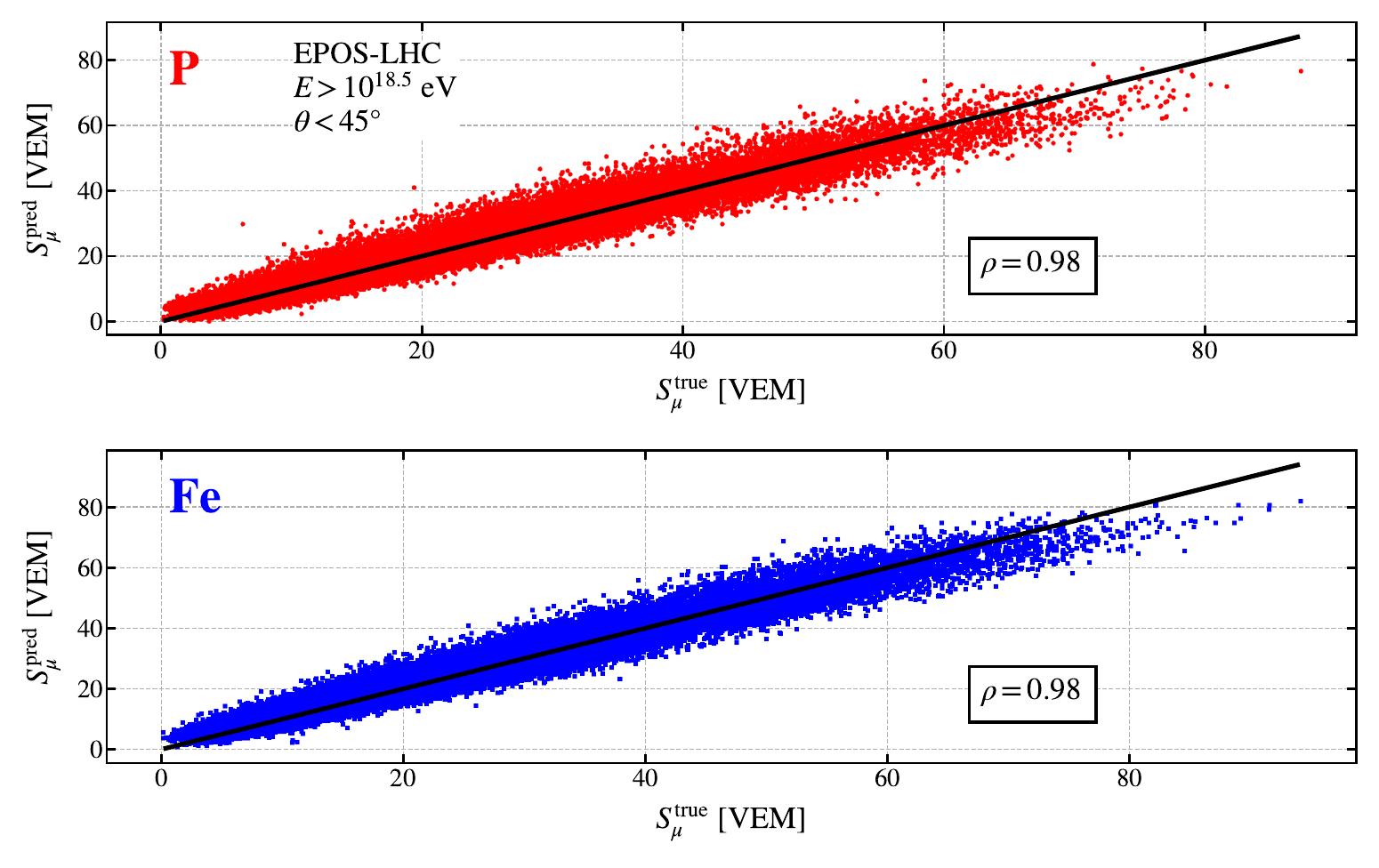}
\caption{Correlation between the true muon signal and the predicted muon signal. Events have been generated using \ep to model hadronic interactions.}
\label{fig:ESmcorr}
\end{figure*}

\begin{figure*}[ht]
\centering
\includegraphics[width=\textwidth]{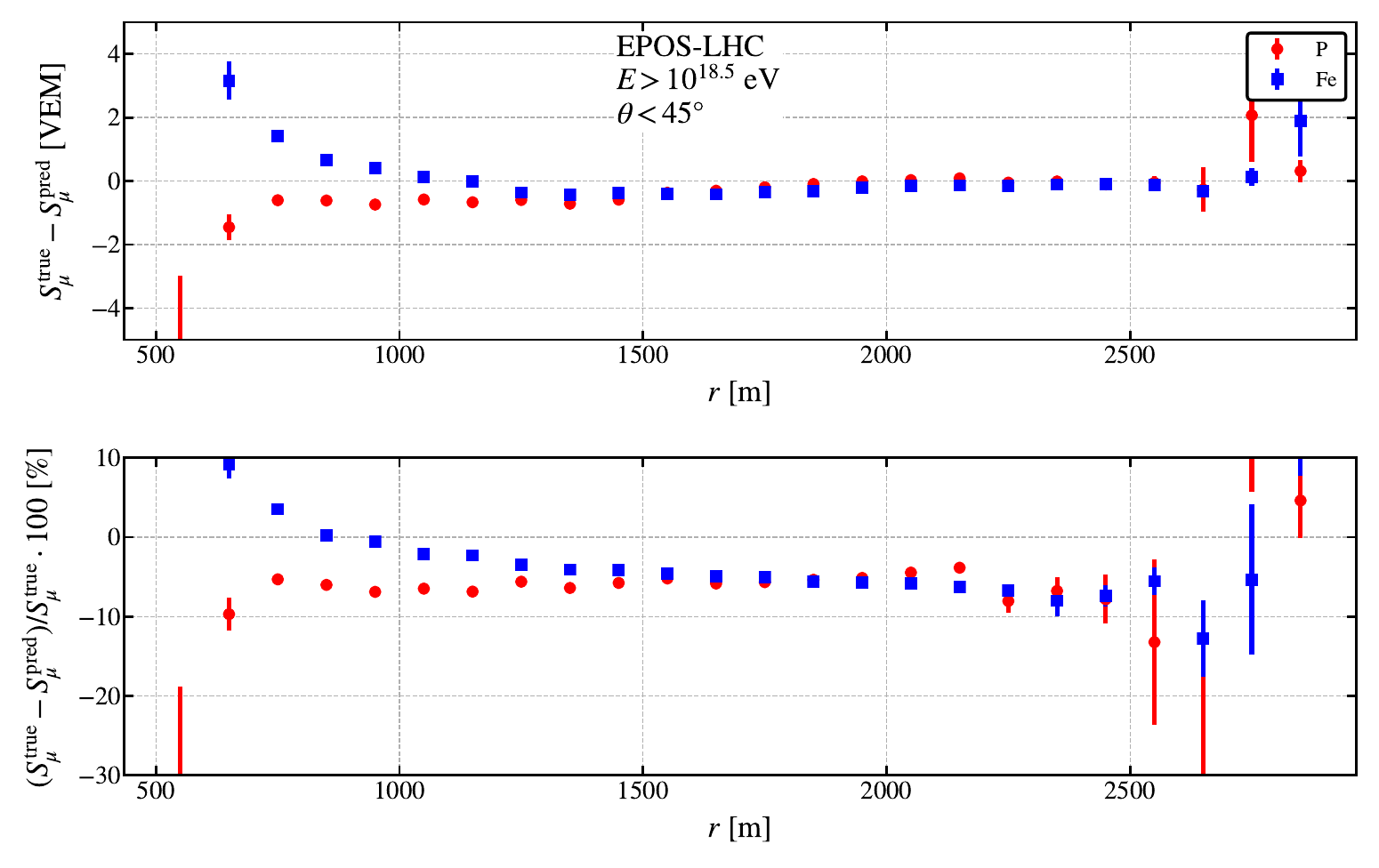}
\caption{Mean of the distribution of differences between true and predicted muon signals as a function of the distance to the core (top panel) and its associated relative error (bottom panel). Events have been generated using \ep to model hadronic interactions.}
\label{fig:ESmr}
\end{figure*}

\begin{figure*}[ht]
\centering
\includegraphics[width=\textwidth]{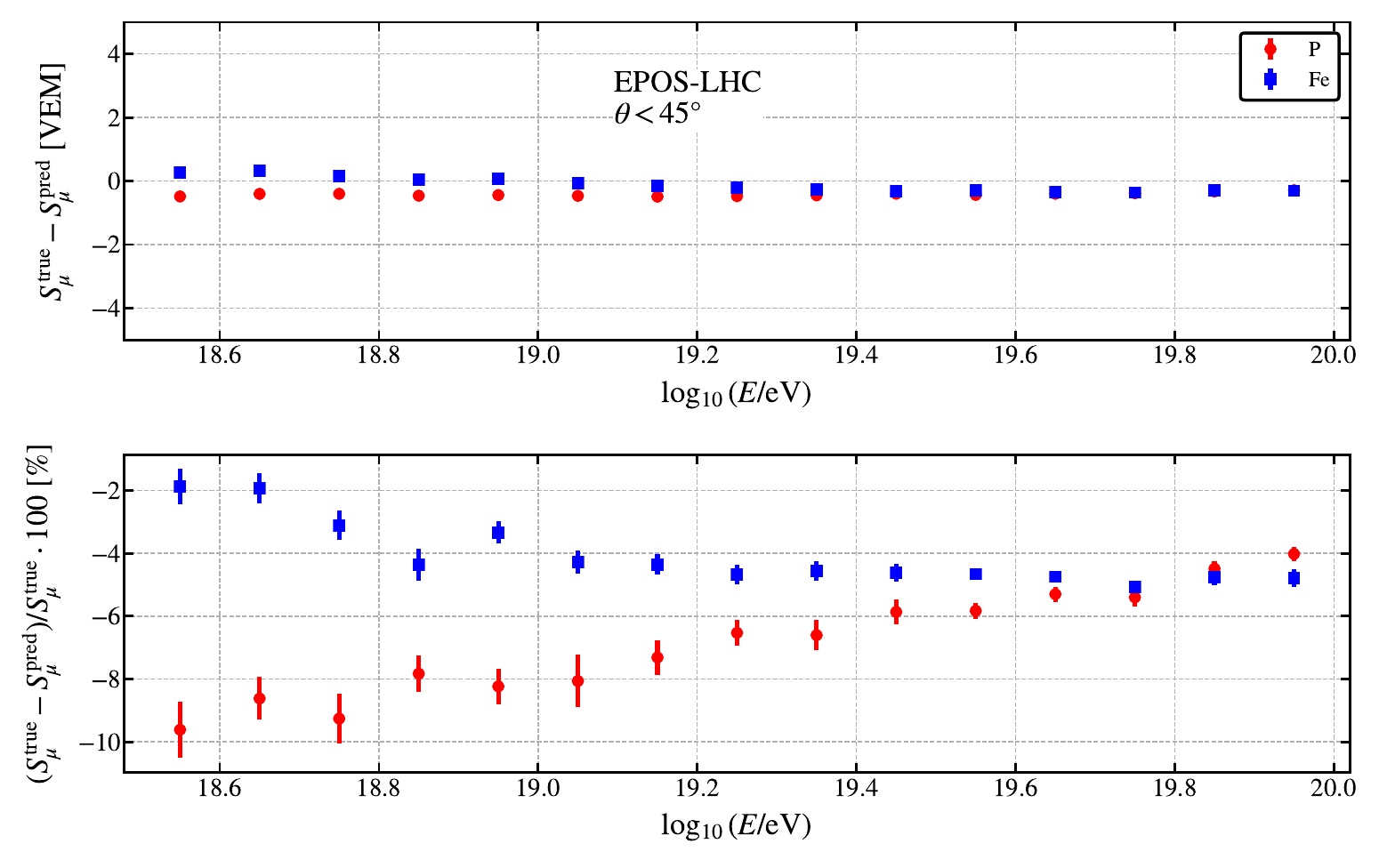}
\caption{Mean of the distribution of differences between true and predicted muon signals as a function of the Monte Carlo event energy (top panel) and its associated relative error (bottom panel). Events have been generated using \ep to model hadronic interactions.}
\label{fig:ESme}
\end{figure*}

\begin{figure*}[ht]
\centering
\includegraphics[width=\textwidth]{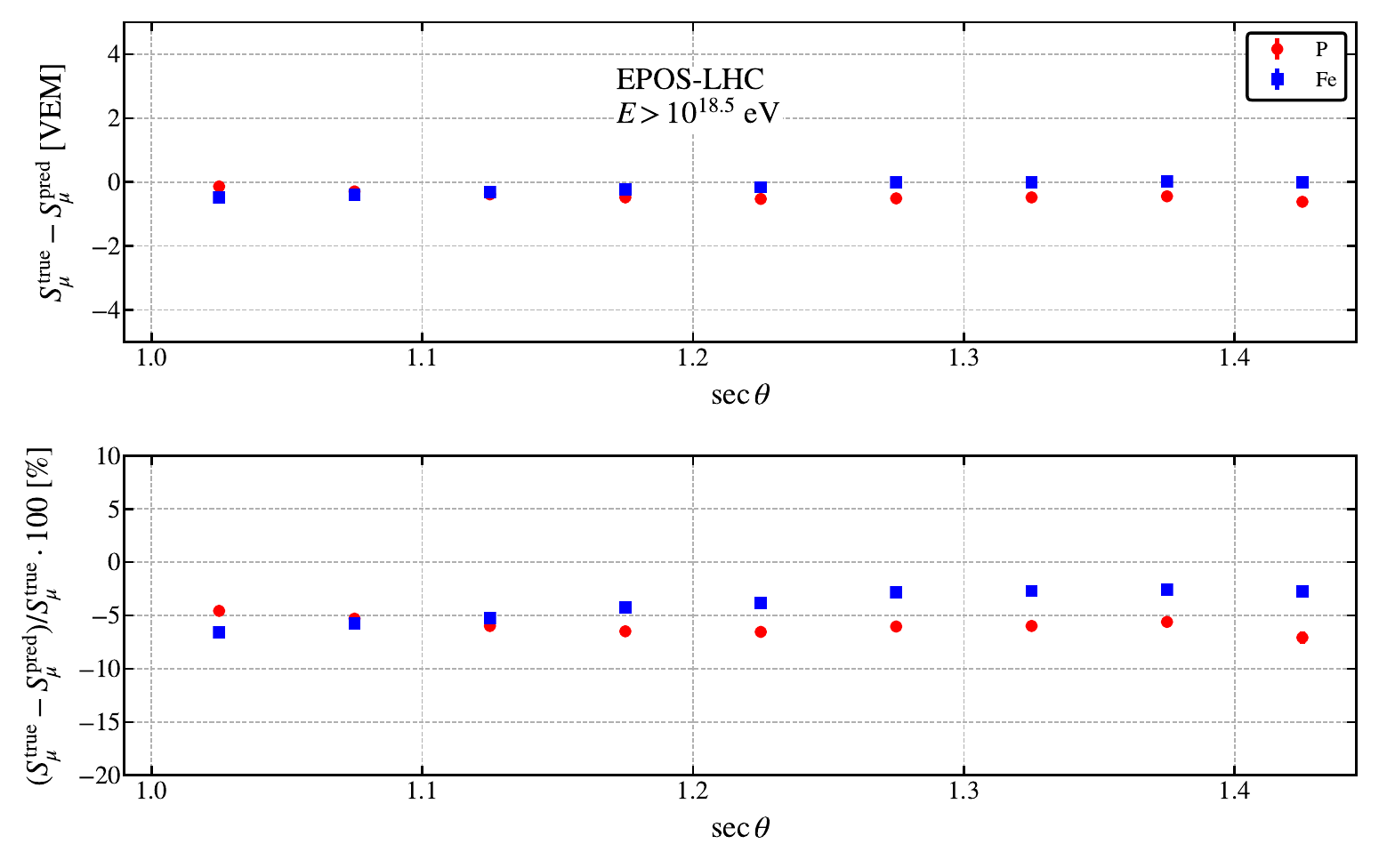}
\caption{Mean of the distribution of differences between true and predicted muon signals as a function of $\sec\theta$ (top panel) and its associated relative error (bottom panel). Events have been generated using \ep to model hadronic interactions.}
\label{fig:ESmt}
\end{figure*}

\begin{figure*}[ht]
\centering
\includegraphics[width=\textwidth]{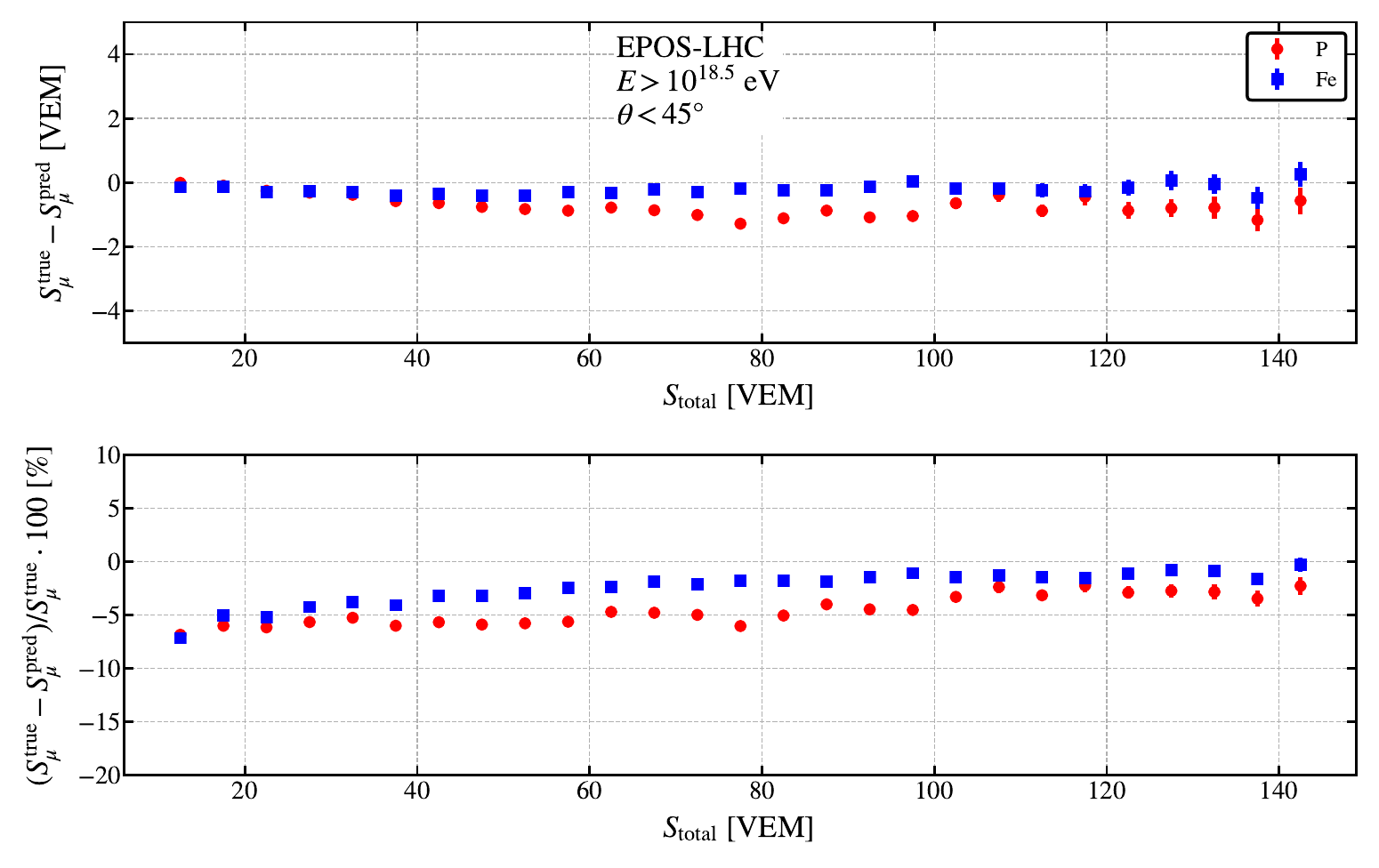}
\caption{Mean of the distribution of differences between true and predicted muon signals as a function of the total signal registered in individual stations (top panel) and its associated relative error (bottom panel). Events have been generated using \ep to model hadronic interactions.}
\label{fig:ESmSt}
\end{figure*}

In our view, a crucial test our approach must overcome is to prove that, once the learning process has finished and the internal architecture of the net has been fixed using events generated with \qgj, the capability to accurately estimate the muon signal in a detector is independent of the hadronic model used. With this goal in mind, we generated a sample of events that used \ep as the model for hadronic interactions (see Table \ref{tab:events}). The results of this exercise are shown in Figures \ref{fig:ESmd}$-$\ref{fig:ESmSt}. These Figures illustrate the robustness of the final DNN to a change in the hadronic models used for testing its performance. The relative error stays below 10\%, and the absolute difference between predicted and true signal does not exceed 2 VEM units. In addition, no sensible bias occurs as a function of the usual variables previously checked. We interpret this as a sign that the correlations between the variables used in the models under consideration are similar and therefore show a high degree of universality.    

\section{Conclusions}
\label{sec:con}

This work explores how machine learning algorithms help in advancing our understanding of the physics of extensive air showers. Taking as an example the Surface Detector of the Pierre Auger Observatory, we have demonstrated that deep learning methods are a powerful tool to estimate the muon signal fraction in the traces registered by the water-Cherenkov detectors of the Auger Surface Detector Array. Based on Monte Carlo studies, we prove that, for each individual station, we obtain accuracies that are typically better than 10\%. 
Our method can be applied to a wide spectrum of primary nuclei, energies, zenith angles and distance ranges. It would be interesting to see how those algorithms perform when applied to experimental data and how they compare with previous studies that reported a substantial discrepancy in the number of muons measured in data and those predicted by simulations. Combined with the future data provided by AugerPrime, machine learning techniques seem a promising tool to gain further insight into the mysteries of ultra-high-energy cosmic rays.    

\newpage
\section*{Acknowledgments}
We warmly thank our colleagues of the Pierre Auger Observatory for their support and for letting us use the official software of the collaboration. We sincerely acknowledge many illuminating discussions with M. Erdmann, D. Veberi\u{c}, A. A. Watson and A. Yushkov. 

This work was done under the auspices of MINECO project FPA2015-70420-C2-2-R. C.J. Todero Peixoto's research was developed with the support of the CENAPAD-SP (Centro Nacional de Processamento de Alto Desempenho em S\~ao Paulo), project UNICAMP / FINEP - MCT.  He also received support through process number: 2016/19764-9, Funda\c{c}\~ao de Amparo \`a Pesquisa do Estado de S\~ao Paulo (FAPESP). J.M. Carceller acknowledges a \textit{Beca de Iniciaci\'on a la Investigaci\'on} fellowship from the UGR.

The authors acknowledge the National Laboratory for Scientific Computing (LNCC/MCTI, Brazil) for providing HPC resources of the SDumont supercomputer, which have contributed to the research results reported in this paper 
(http://sdumont.lncc.brs).

\bibliographystyle{elsarticle-num}
\bibliography{References}

\end{document}

%% file: Fig/genetic-algorithm.pdf_tex
\begingroup%
  \makeatletter%
  \providecommand\color[2][]{%
    \errmessage{(Inkscape) Color is used for the text in Inkscape, but the package 'color.sty' is not loaded}%
    \renewcommand\color[2][]{}%
  }%
  \providecommand\transparent[1]{%
    \errmessage{(Inkscape) Transparency is used (non-zero) for the text in Inkscape, but the package 'transparent.sty' is not loaded}%
    \renewcommand\transparent[1]{}%
  }%
  \providecommand\rotatebox[2]{#2}%
  \newcommand*\fsize{\dimexpr\f@size pt\relax}%
  \newcommand*\lineheight[1]{\fontsize{\fsize}{#1\fsize}\selectfont}%
  \ifx\svgwidth\undefined%
    \setlength{\unitlength}{651.08267212bp}%
    \ifx\svgscale\undefined%
      \relax%
    \else%
      \setlength{\unitlength}{\unitlength * \real{\svgscale}}%
    \fi%
  \else%
    \setlength{\unitlength}{\svgwidth}%
  \fi%
  \global\let\svgwidth\undefined%
  \global\let\svgscale\undefined%
  \makeatother%
  \begin{picture}(1,0.69387753)%
    \lineheight{1}%
    \setlength\tabcolsep{0pt}%
    \put(0,0){\includegraphics[width=\unitlength,page=1]{./Fig/genetic-algorithm.pdf}}%
    \put(0.79173709,0.64569625){\color[rgb]{0,0,0}\makebox(0,0)[t]{\lineheight{0}\smash{\begin{tabular}[t]{c}selection\end{tabular}}}}%
    \put(0.79030449,0.01334718){\color[rgb]{0,0,0}\makebox(0,0)[t]{\lineheight{0}\smash{\begin{tabular}[t]{c}crossover\end{tabular}}}}%
    \put(0.18670154,0.0133978){\color[rgb]{0,0,0}\makebox(0,0)[t]{\lineheight{0}\smash{\begin{tabular}[t]{c}mutation\end{tabular}}}}%
    \put(0,0){\includegraphics[width=\unitlength,page=2]{./Fig/genetic-algorithm.pdf}}%
    \put(0.18670154,0.64569625){\color[rgb]{0,0,0}\makebox(0,0)[t]{\lineheight{0}\smash{\begin{tabular}[t]{c}evaluation\end{tabular}}}}%
    \put(0,0){\includegraphics[width=\unitlength,page=3]{./Fig/genetic-algorithm.pdf}}%
  \end{picture}%
\endgroup%